\documentclass[preprint2]{emulateapj}

\usepackage{graphicx}
\usepackage{color}
\usepackage{array}
\usepackage{rotating}
\usepackage{amsmath}
\usepackage{amssymb}
\usepackage{lineno}
\usepackage{appendix}

\newif\ifarxiv

\arxivtrue
\ifarxiv
\else
\linenumbers
\fi

\usepackage[version=3]{mhchem} % Package for chemical equation typesetting

\def\E{\mathrm{e}}
\def\erf{\mathrm{erf}}

\definecolor{blueish}{rgb}{0.2, 0.8, 0.8}

\shorttitle{Stochastic \ce{CO2} and Snowball Glaciation}
\shortauthors{Wordsworth}

\begin{document}

%% LaTeX will automatically break titles if they run longer than
%% one line. However, you may use \\ to force a line break if
%% you desire.

\title{How likely are Snowball episodes near the inner edge of the habitable zone?}

%% Use \author, \affil, and the \and command to format
%% author and affiliation information.
%% Note that \email has replaced the old \authoremail command
%% from AASTeX v4.0. You can use \email to mark an email address
%% anywhere in the paper, not just in the front matter.
%% As in the title, use \\ to force line breaks.

\author{R.~Wordsworth}
\affil{School of Engineering and Applied Sciences, Harvard, Cambridge, MA 02138, USA}
\affil{Department of Earth and Planetary Sciences, Harvard, Cambridge, MA 02138, USA}
\email{rwordsworth@seas.harvard.edu}

%% Keywords should appear after the \end{abstract} command. The uncommented
%% example has been keyed in ApJ style. See the instructions to authors
%% for the journal to which you are submitting your paper to determine
%% what keyword punctuation is appropriate.


\begin{abstract}
Understanding when global glaciations occur on Earth-like planets is a major challenge in climate evolution research. Most models of how greenhouse gases like \ce{CO2} evolve with time on terrestrial planets are deterministic, but the complex, nonlinear nature of Earth's climate history motivates study of non-deterministic climate models. Here a maximally simple stochastic model of \ce{CO2} evolution and climate on an Earth-like planet with an imperfect \ce{CO2} thermostat is investigated. It is shown that as stellar luminosity is increased in this model, the decrease in the average atmospheric \ce{CO2} concentration renders the climate increasingly unstable, with excursions to a low-temperature state common once the received stellar flux approaches that of present-day Earth. Unless climate feedbacks always force the variance in \ce{CO2} concentration to decline rapidly with received stellar flux, this means that terrestrial planets near the inner edge of the habitable zone may enter Snowball states quite frequently. Observations of the albedos and color variation of terrestrial-type exoplanets should allow this prediction to be tested directly in the future.
\end{abstract}
 
\maketitle

Investigating the processes that determine planetary habitability and predicting their observable consequences is a key objective of exoplanet climate modeling \citep{seager2013exoplanet}. Today, Earth is still our only confirmed example of a habitable planet, so its climate and chemistry continues to drive our understanding of habitability in general.  One of the most influential models of long-term \ce{CO2} evolution on Earth is the carbonate-silicate weathering feedback \citep{walker1981negative}, which is the basis for the `canonical' definition of the habitable zone \citep{kasting1993habitable,kopparapu2013habitable}. Despite the popularity of this model, the nature of Earth's \ce{CO2} cycle through geologic time remains a highly active area of research, and a number of processes likely cause Earth's carbon cycle to deviate significantly from the standard weathering feedback \citep{maher2014hydrologic,macdonald2019arc,graham2020thermodynamic}.

Accurate estimates of temperature and atmospheric \ce{CO2} in Earth's deep-time history are difficult to obtain, but there is no evidence for secular warming of the climate over the last 4~Gy \citep{feulner2012faint}. Because the Sun's luminosity has increased with time (by about 30 to 40\% over the last 4~Gy) and \ce{CO2} has likely been a key greenhouse gas throughout Earth history, a secular decline in atmospheric \ce{CO2} with time seems almost certain. However, this decline has been far from monotonic: current anthropogenic emissions aside, the variations in Earth's surface temperature and atmospheric \ce{CO2} levels just in the last 400~My have been substantial, for reasons that are still the subject of intensive study \citep{franks2014new,montanez2016climate,lenardic2016climate,macdonald2019arc}. 

Motivated by these observations and previous modeling efforts, the purpose of this note is to construct a simple stochastic model of \ce{CO2} evolution, and to apply it to terrestrial-type planets. The model is intentionally semi-empirical, rather than mechanistic, because many of the processes that affect Earth's \ce{CO2} levels remain so uncertain. As will be shown, the transition to a stochastic view of \ce{CO2} evolution leads to qualitatively different conclusions compared to the deterministic picture.  

Surface temperature evolution in the model is represented as
\begin{equation}
C\frac{dT}{dt} =  \frac 14 F [1-A(T)] - OLR  \label{eq:ebal}
\end{equation}
where $T$ is surface temperature, $C$ is the heat capacity of the ocean-atmosphere system (here in J/m$^2$/K), $F$ is incident stellar flux, and $OLR$ is the outgoing longwave radiation at the top of the atmosphere, which we will take to be a function of $T$ and the molar concentration of \ce{CO2} in the atmosphere.  Internal climate variability, which would add a stochastic term to \eqref{eq:ebal}, is neglected here to keep the focus on the impact of variability in the \ce{CO2} cycle. 

The aim here is to point out general model features rather than to make precise predictions, so we linearize the OLR around Earth's preindustrial surface temperature $T_0=288$~K and \ce{CO2} molar concentration \mbox{$f_{\ce{CO2},0}=280$~ppmv}:
\begin{equation}
OLR \approx  OLR_0 + a(T-T_0) - b\log (f_{\ce{CO2}}/f_{\ce{CO2},0}). \label{eq:OLRsimple}
\end{equation}
Here $a=2$~W/m$^2$/K following \cite{abbot2016analytical}, and $b = 5.35$~W/m$^2$ is the radiative forcing coefficient for \ce{CO2} \citep{myhre1998new}. Logarithmic dependence of OLR on $f_{\ce{CO2}}$ is a reasonable approximation in the 10 to $10^5$~ppmv \ce{CO2} and 280 to 290~K temperature range, although the value of $b$ begins to increase at high \ce{CO2} concentrations.

Setting $F = F_0 + \Delta F$ and noting that \mbox{$\frac 14 F_0(1-A_0) = S_0 = OLR_0$}, where $F_0=1366$~W/m$^2$ and $A_0=0.3$ are Earth's present-day received solar flux and albedo, respectively, we can write the time evolution of the temperature deviation from the baseline state $x=T-T_0$ as 
\begin{equation}
C\frac{dx}{dt} = \Delta S - \frac 14 F \Delta A  - ax + b\log y  \label{eq:ebal_linear}
\end{equation}
where $y=f_{\ce{CO2}}/f_{\ce{CO2},0}$, $\Delta A = A(T) - A_0$ and \mbox{$\Delta S  =  \Delta F (1-A_0)/4$} \citep{abbot2016analytical}. $\Delta A$ is quite hard to assess in general due to cloud effects, but its dependence on surface ice coverage acts to accelerate Snowball transitions as the transition temperature is approached. As the main aim here is to assess the likelihood of a Snowball transition as a function of atmospheric \ce{CO2} concentration, rather than to study the Snowball state itself, it can be safely set to zero. In addition, the climate achieves thermal balance far more rapidly than atmospheric \ce{CO2} levels change, so we set ${dx}/{dt}=0$. This allows the temperature deviation from the present-day Earth value to be written as
\begin{equation}
x = \frac{ \Delta S + b\log y }{a} \label{eq:x_vs_y}.
\end{equation}

Next, we incorporate \ce{CO2} evolution. The evolution of the \ce{CO2} molar concentration $y$ with time is modeled as an Ornstein-Uhlenbeck process with an offset term $\chi$ \citep{jacobs2010stochastic}. For a given timestep $dt$, this means that the increment in $y$ is 
\begin{equation}
dy = -\tau^{-1} (y-\chi) dt + g dW \label{eq:main_stoch}
\end{equation}
Here $g$ is a constant and $dW$ represents a Wiener process such that for every timestep $dt$, $dW$ is equal to a value taken from a gaussian distribution with variance $dt$. $\tau$ is a timescale that determines how rapidly $y$ is drawn back to the mean value (either by carbonate-silicate weathering feedbacks, or some other process). At each timestep, $y$ is set to $-y$ if $y<0$, ensuring that $y$ always remains positive-valued.

Equation~\eqref{eq:main_stoch} provides an inherently non-deterministic representation of \ce{CO2} evolution, with a linear restoration term that prevents unbounded growth in the probability distribution for $y$ with time. From any starting condition, the system evolves towards a statistically steady state on a timescale $\tau$. Once a steady state is reached, the probability density function for $y$ has the form
\begin{equation}
q(y) = Q\E^{-(y-\chi)^2/2\sigma_y^2} \label{eq:y_pdf}
\end{equation}
where the standard deviation $\sigma_y =\sqrt{\tau/2} g$ and the normalization factor
\begin{equation}
Q = \frac{2}{\sigma_y\sqrt{2\pi}  }  \frac 1{1 + \erf[\chi/\sqrt 2 \sigma_y]} 
\end{equation}
ensure
\begin{equation}
\int_0^\infty q(y) dy = 1.
\end{equation}
Because $y$ can only take positive values, the mean of this distribution is
\begin{equation}
\overline y =\int_0^\infty y q(y) dy = \chi + \sigma_y^2q_0.
\end{equation}
where $q_0 = q(0)$.  The distribution variance is
\begin{equation}
V = \int_0^\infty (y - \overline y)^2 q(y) dy = \sigma_y^2(1 - \overline y q_0).
\end{equation}
Note that when $\chi >> \sigma_y$, $\overline y = \chi$ and $V = \sigma_y^2$. 

Converting \eqref{eq:main_stoch} into a statement about the probability of a Snowball transition under given conditions requires the parameter $\chi$ to be determined. Given the lack of evidence for secular warming of Earth over the last 4~Gy as the Sun's luminosity has increased, the simplest assumption we can make is that \ce{CO2} feedbacks set $\chi$ at a value that yields $\overline T = T_0$ (and hence $\overline x=0$) on long timescales. Taking the time mean of \eqref{eq:ebal_linear} and assuming separation of timescales between slow ($>100$~My) evolution of $\Delta S$ and more rapid stochastic \ce{CO2} fluctuations yields
\begin{equation}
\overline{\log y} = -\Delta S/b = S_0\left(1 - \alpha \right)/b \label{eq:x_vs_y_av}
\end{equation}
where $\alpha = F/F_0$ is the stellar flux received relative to Earth's present-day received flux. $\chi$ is then calculated for use in \eqref{eq:main_stoch} by finding the root of the function
\begin{equation}
\Phi(\chi) = \overline {\log y} -  \int_{0}^{\infty} \log y q(y,\chi) dy \label{eq:chi_rootfind}
\end{equation}
numerically. Situations where $\Phi(\chi)$ has no root for $\chi>0$ occur at low $\chi/\sigma_y$ values, but this is of little practical significance here, because the planet enters a Snowball state before they are reached.

\begin{figure}[h]
	\begin{center}
		{\includegraphics[width=5in]{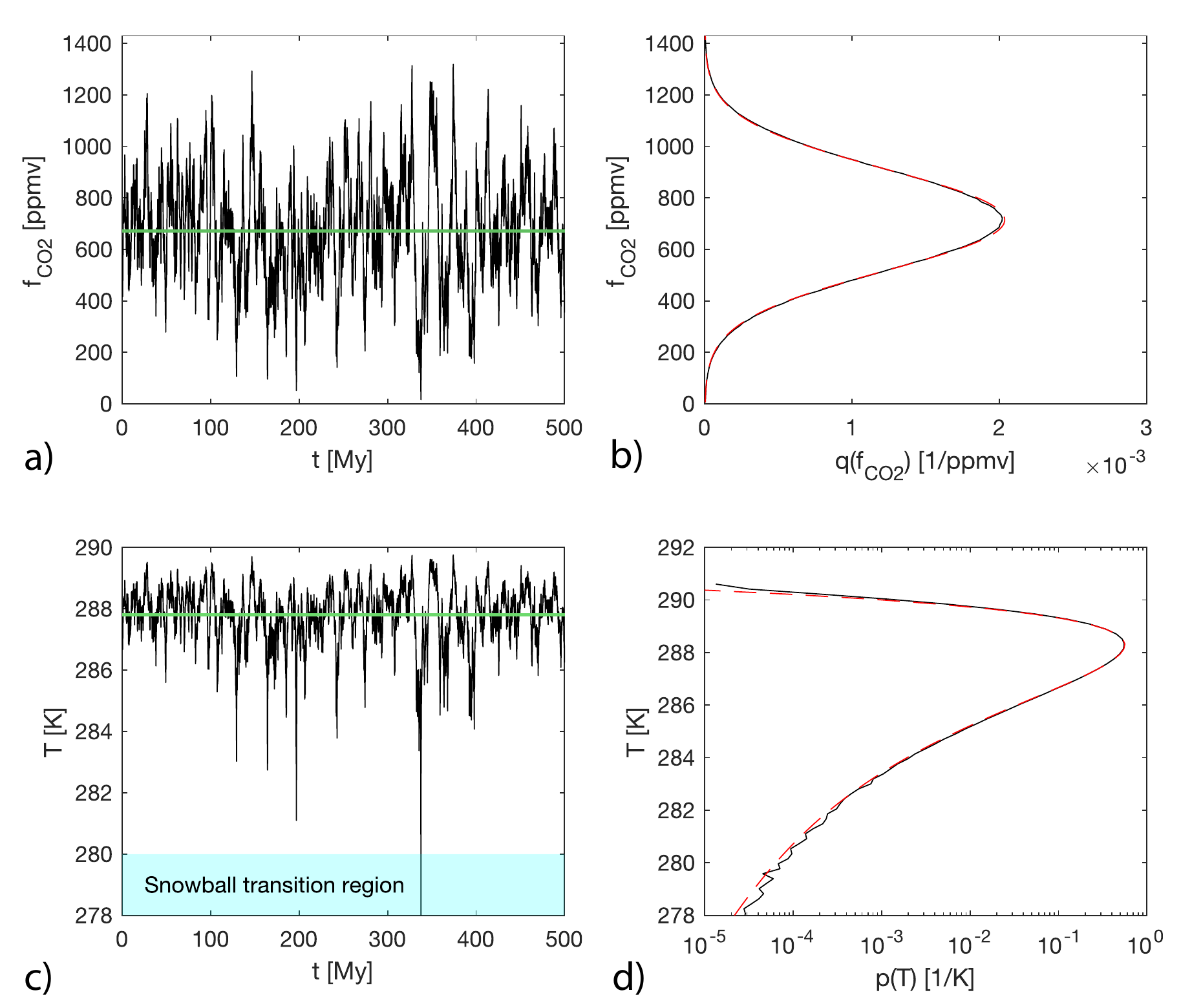}}
	\end{center}
	\caption{Output of the numerical stochastic model over 500~My for an ensemble of 1024 runs, given fixed stellar luminosity ($\alpha = 0.98$), relaxation timescale $\tau = 2.5$~My, and $\sigma_y = 0.7$. a) \ce{CO2} molar concentration vs. time, b) normalized histogram of \ce{CO2} molar concentration values, c) temperature vs. time and d) normalized histogram of temperature values. For a) and c), a single run where temperature dropped below the Snowball limit is shown in black. In c) the light blue shading indicates the Snowball transition region, while in both a) and c), the green line indicates the mean value. Finally, in b) and d) the solid black and red dashed lines indicate the numerical results and analytic results according to \eqref{eq:y_pdf} and \eqref{eq:x_pdf}, respectively. The ice-albedo feedback is not included in these simulations.}
\label{fig:prob}
\end{figure}

Because temperature depends directly on the \ce{CO2} concentration, a second probability density function $p(x)$ for the temperature deviation $x$ can be written as 
\begin{equation}
p(x) = q[y(x)]\left| \frac{dy}{dx} \right|.\label{eq:prob_defn}
\end{equation}
\newline
We can rearrange \eqref{eq:x_vs_y} in terms of $y$, take the derivative, and substitute in the result along with \eqref{eq:y_pdf} to get 
\begin{equation}
p(x) = c Q \E^{c x-\Delta S /b - \frac 12(\E^{c x-\Delta S/b}-\chi)^2/\sigma_y^2} .  \label{eq:x_pdf}
\end{equation}
where $c = a/b$. Given \eqref{eq:prob_defn}, we can also write \mbox{$q(f_{\ce{CO2}})=q(y)/f_{\ce{CO2},0}$} and $p(T)=p(x)$. The function $p(x)$ is asymmetric, with rapid decline at high $x$ values but a long tail stretching to low $x$ values (Fig.~\ref{fig:prob}). The implication is that for a gaussian \ce{CO2} concentration distribution with a given variance, very low temperatures are reached more frequently than very high temperatures. Hence even when the mean \ce{CO2} concentration is well above the threshold for a Snowball event, there remains a finite probability of a transition occurring.

Figure~\ref{fig:prob} shows the results of solving \eqref{eq:main_stoch} numerically via the Euler method over 1~Gy of constant stellar luminosity. \ce{CO2} and temperature time series are shown for a single run (panels a and c), and probability density functions $q(f_{\ce{CO2}})$ and $p(T)$ are shown for an ensemble of 1024 runs (panels b and d). The asymmetry of the temperature evolution indicated by \eqref{eq:x_pdf} is clear from Fig.~\ref{fig:prob}d. In the time series shown, \ce{CO2} levels temporarily dip low enough to make temperature fall below the Snowball threshold \citep[set here at 280~K, following][]{pierrehumbert2011climate} just after 300~My. This was verified to cause a Snowball transition when ice-albedo effects were included in the model (results not shown). 

\begin{figure}[h]
	\begin{center}
		{\includegraphics[width=3.5in]{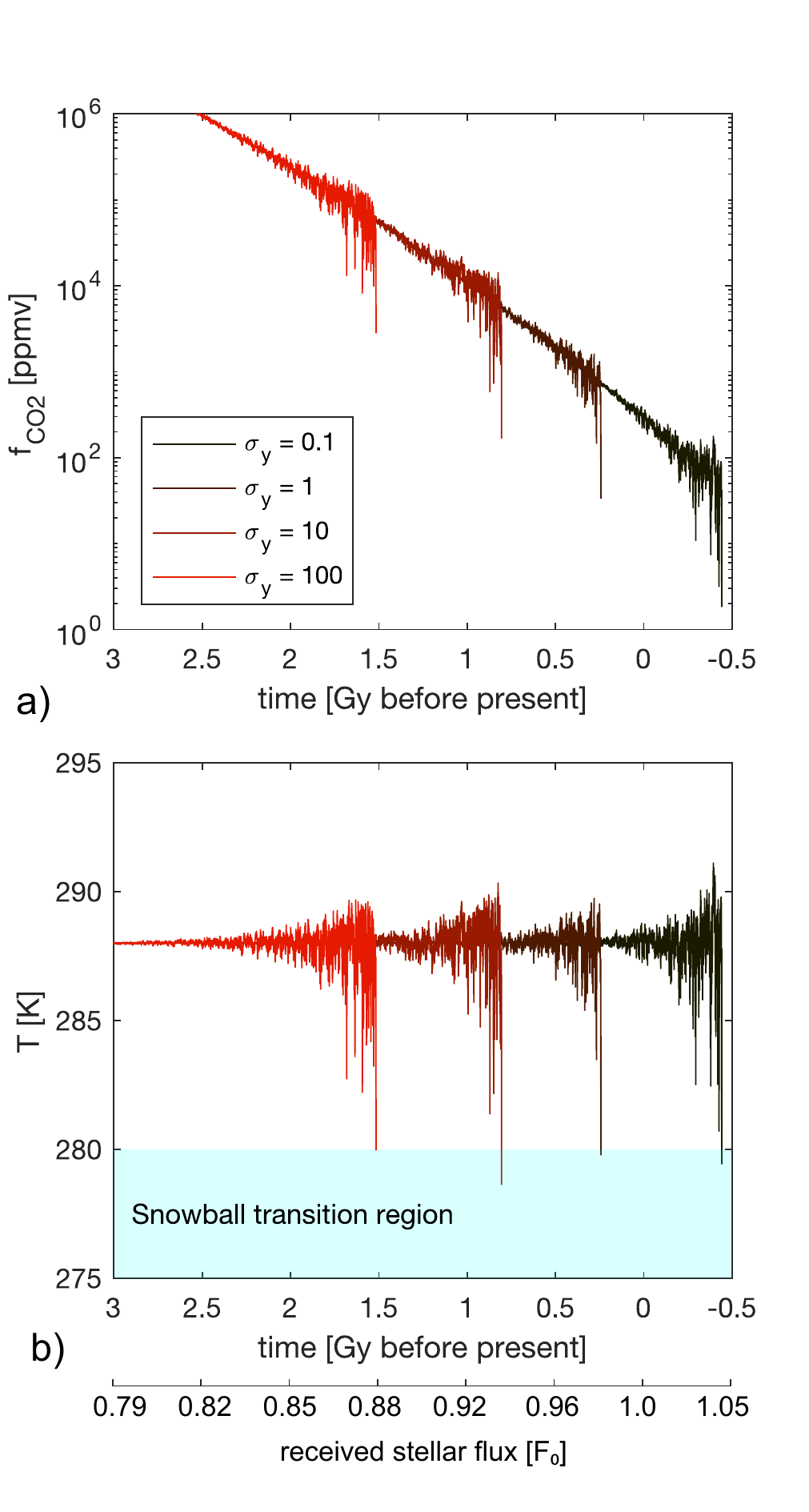}}
	\end{center}
	\caption{Output of the numerical stochastic model for an Earth-like planet around a G-star with evolving stellar luminosity over 3.5~Gy, starting from 3 Gy before present. a) Atmospheric \ce{CO2} molar concentration and b) global mean temperature as a function of time. Received stellar flux is also shown on the $x$-axis. In b), the light blue shading indicates the Snowball transition region.}
\label{fig:evol}
\end{figure}

Figure~\ref{fig:evol} shows the output of the same model when secular evolution of stellar flux is included. Here, 3.5~Gy of evolution is simulated and stellar flux evolution is represented as
\begin{equation}
\alpha(t) = \frac{1}{1 + \frac 25(1-t/{4.5~\mbox{Gy}})}
\end{equation}
which is appropriate for a Sun-like star \citep{gough1981solar}. Generalization of the results to other star types is straightforward in principle, but it is not pursued here, in part because Snowball transitions on low-mass stars may be strongly affected by the stellar spectrum and the planet's spin-orbit configuration \citep{joshi2012suppression,shields2013effect,checlair2017no}.

As can be seen from Fig.~\ref{fig:evol}, for a fixed value of $\sigma_y$, temperature fluctuations steadily increase with time until a Snowball transition occurs, with larger $\sigma_y$ values yielding earlier Snowball transitions. A Snowball event at some point in time is therefore inevitable unless $\sigma_y$ declines at least as fast as $\chi$ does. This result shows that if a planet possess an effective \ce{CO2} thermostat on long timescales ($>100$~My), as Earth appears to, but the shorter term variance in \ce{CO2} does not decline rapidly as stellar luminosity increases, the chance of undergoing a Snowball glaciation should \emph{increase} as the planet gets closer to the inner edge of the habitable zone. This is a very different prediction from that of deterministic models of \ce{CO2} cycles on Earth-like planets, which either predict permanently clement conditions, or glaciations that only begin to occur towards the outer edge of the habitable zone \citep[e.g., ][]{tajika2007long,haqq2016limit}.

\begin{figure}[h]
	\begin{center}
		{\includegraphics[width=3.5in]{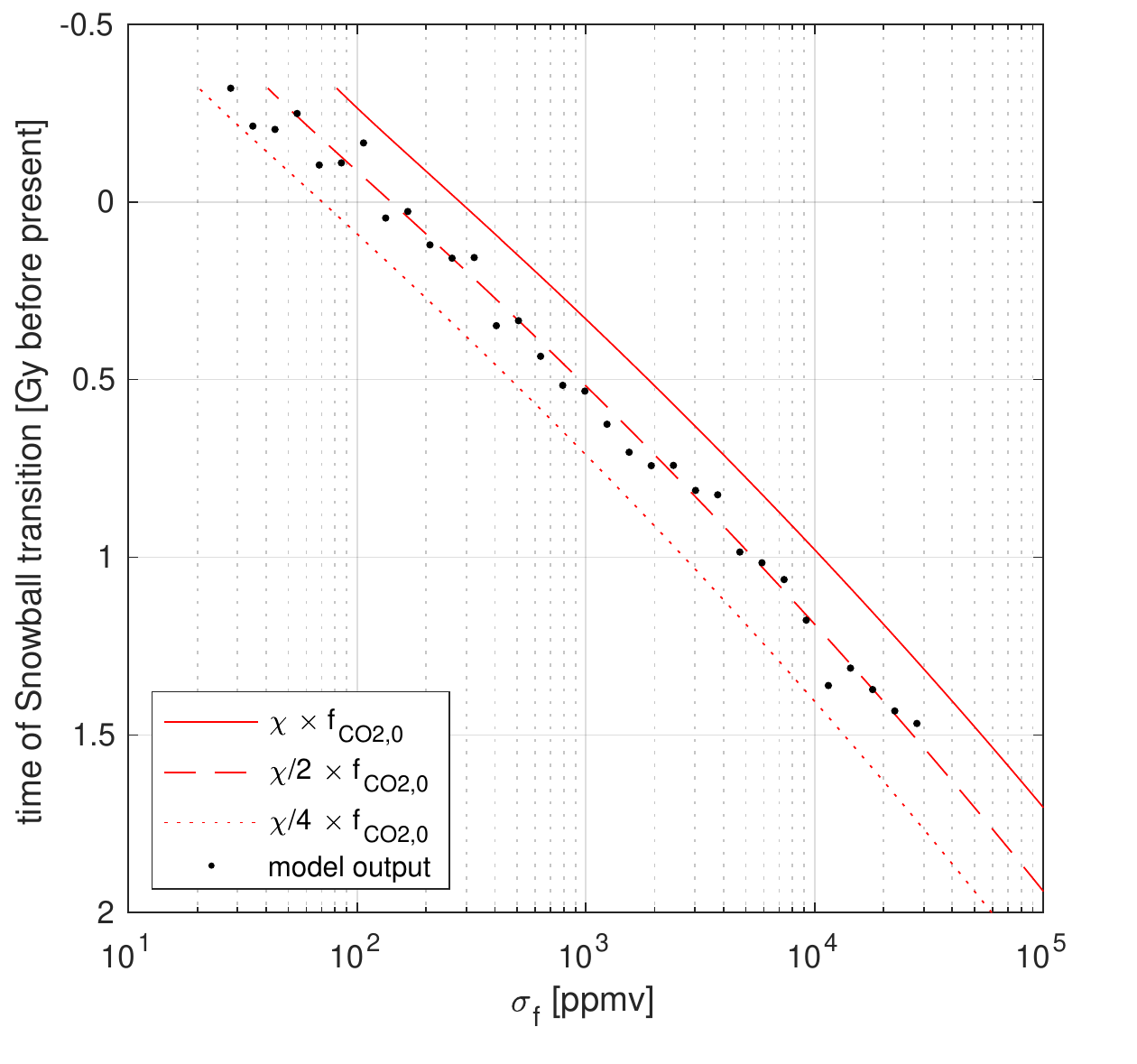}}
	\end{center}
	\caption{Time of Snowball transition as a function of \ce{CO2} concentration standard deviation $\sigma_f$ in ppmv ($\sigma_f = \sigma_y \times 280$~ppmv). Black dots show numerical model results, while the red lines show fractions of the steady-state parameter $\chi$, which is calculated from \eqref{eq:chi_rootfind}.}
\label{fig:snowball_trans}
\end{figure}

Figure~\ref{fig:snowball_trans} shows the time of Snowball transition for 32 simulations with different values of $\sigma_y$. There is some scatter because of the stochastic nature of the simulations, but the strong dependence of transition time on the \ce{CO2} variance is clear. As a general rule, once $\chi$ drops to below 2 to 4 times the value of $\sigma_y$, a Snowball transition becomes likely. The effect of $\tau$ on the results was also tested, and it was found that a larger $\tau$ caused transitions to occur at a given time at slightly higher $\sigma_y/\chi$ ratios in general, although the effect was not large in the $\tau = 0.1$ to $10$~My range.

Short-term climate variability due to effects like stellar fluctuations, ocean-atmosphere feedbacks and volcanic aerosol emissions \citep[e.g., ][]{foster2011global,macdonald2017initiation,arnscheidt2020routes} has been neglected here. Recent work has elegantly demonstrated the impact of short-term climate variability on Snowball transitions in a probabilistic framework \citep{lucarini2019transitions}. Including this variability would simply increase the probability of a Snowball transition for a given set of parameters in equation~\eqref{eq:main_stoch}. Of course, on a planet where stellar, albedo and dynamical variability always dominates variations in greenhouse gas concentrations, the trend described above would be masked by these effects. However, for Earth at least it is clear that variations in \ce{CO2} concentration have been a fundamental driver of climate change over geologic time.

Reliable constraints on $\sigma_y$ on Earth on long timescales are hard to come by, although it does appear to have declined over the last Gy or so since the Neoproterozoic Snowball events. Three-dimensional climate modeling suggests the Marinoan Snowball glaciation that terminated 635~My ago occurred at a \ce{CO2} concentration of between 280 and 560~ppmv \citep{voigt2011initiation,voigt2012sea}. This can be compared with the 3000~ppmv or more \ce{CO2} that would likely have been present under temperate or warm climate conditions \citep{pierrehumbert2011climate}. Over the last 400~My, the characteristics of fossil leaf stomata and other proxies constrain \ce{CO2} to around 250--2000~ppmv, with the greatest uncertainty at the highest concentrations \citep{franks2014new}.  Finally, over just the last 800~ky until the industrial era, \ce{CO2} has varied between about 280 and 180~ppmv \citep{luethi2008high}, with a standard deviation $\sigma_y$ of $0.11$ of the mean value (mean {$f_{\ce{CO2}} = 224$~ppmv}, {$\sigma_f = 25$~ppmv}).  A few hundred My in the future, the stochastic model predicts that \ce{CO2} fluctuations of this order ($\sigma_f$ of a few 10s of ppmv) would be sufficient to start a Snowball transition (Fig.~\ref{fig:snowball_trans}).

The model presented here is extremely simple and empirical. However, it is arguably at least as justified for exoplanet habitability modeling as the many other more sophisticated deterministic models of \ce{CO2} evolution that currently exist. It is of course possible that variance in $f_{\ce{CO2}}$ does always decrease rapidly enough as stellar luminosity increases to prevent Snowball transitions. However, even if \ce{CO2} variance has decreased on Earth since the Neoproterozoic, it is not at all obvious based on our current understanding of the carbon cycle that this trend will continue to hold in the future, or apply in general to Earth-like exoplanets.

The long-term \ce{CO2} source in the carbon cycle (volcanism) behaves largely independently of climate until Venus-like atmospheric pressures are reached, with possible modest \emph{positive} feedbacks due to couplings between sea level and the rate of mid-ocean ridge volcanism \citep{huybers2017delayed}. The \ce{CO2} weathering sink has a temperature dependence, but this can readily become saturated because of local physical weathering rate limits, even when the global climate is temperate. Earth's history in the Phanerozoic  \citep{maher2014hydrologic,macdonald2019arc} and Neoproterozoic \citep{hoffman1998neoproterozoic} indicates that the effects of tectonic processes and continental drift on \ce{CO2} variability are both extremely important. Large Igneous Province (LIP) eruptions, which have appeared intermittently throughout Earth history, are capable of supplying huge quantities of weatherable basalt to the surface and hence drawing down large quantities of \ce{CO2}, even at the cooler equatorial temperatures expected near a Snowball transition. Indeed, weathering associated with LIPs likely played a major role in the first Neoproterozoic Snowball Earth transition \citep{cox2016continental}.

The final major source of complexity in Earth's \ce{CO2} cycle is the biosphere. It is plausible, although certainly not guaranteed, that life itself can rapidly reduce \ce{CO2} variance as stellar luminosity increases, which would be a clear example of a `Gaian' feedback \citep{lovelock1974atmospheric}. Surface weathering by land plants is an important part of the modern carbon cycle, and it has been suggested that weathering feedbacks involving C$_3$-photosynthetic plants may have buffered minimum $f_{\ce{CO2}}$ values to 100-200 ppmv on Earth over the last 24~My \citep{pagani2009role}. However, positive biogeochemical ocean feedbacks are a plausible explanation for the ice-age \ce{CO2} oscillations observed over the last 800,000~ky \citep{sigman2010polar}, and coal formation in the Carboniferous may have brought Earth closer to a Snowball than at any time since the Neoproterozoic \citep{feulner2017formation}. Even during the last glacial maximum ca. 20,000~years ago, estimates of global mean temperature \citep{tierney2020glacial} suggest only a few Kelvins of additional cooling could have been sufficient to push Earth into a Snowball state. As our current era of anthropogenic global warming makes clear, when the starting atmospheric \ce{CO2} inventory is small, even relatively small changes in exchange rates with other reservoirs in the system have the capacity to cause sudden and dramatic shifts in climate.

Tests of the canonical carbonate-silicate weathering hypothesis via observations of atmospheric \ce{CO2} on exoplanets has been proposed \citep{bean2017statistical}, although they require a large sample size of planetary targets and highly capable observing systems to be successful \citep{lehmer2020carbonate}. Broadband or spectrally resolved albedo measurements to identify planets in a Snowball state provide an alternative approach, at least as long as degeneracies associated with planetary radius and the presence of thick cloud decks can be addressed \citep{cowan2011rotational,guimond2018direct}. 
Such observations would allow a powerful probe into the level of control that climate feedback mechanisms on terrestrial-type planets provide, and the extent to which Earth's climate history has been unusual.

\section*{Acknowledgments}
This article has benefitted from discussions with E. Tziperman and A. Knoll and helpful comments from an anonymous reviewer. Code to reproduce the plots in the paper is available open-source at \emph{https://github.com/wordsworthgroup/stochastic\_snowball\_2021}. Finally, I thank R. Pierrehumbert for bringing a manuscript preprint by R.~J. Graham \citep{graham2021high} to my attention during the review process that independently puts forward ideas related to those presented here, although with a different focus and modeling approach.

\newpage

\end{document}